\documentclass{elsart}
\usepackage{graphicx,amssymb,amsmath}
\usepackage{wrapfig,bm,color,extarrows}

\usepackage{extarrows}	
\begin{document}

\journal{Acta Biotheoretica}
\begin{frontmatter}
\title{Replication via Invalidating the Applicability of the Fixed Point Theorem}
\author{GentaIto}
\ead{cxq02365@gmail.com}
\address{Maruo Lab., 500 El Camino Real \#302, Burlingame, CA 94010, United States.}
\begin{abstract}
We present a construction of a certain infinite complete partial order (CPO) that differs from the standard construction used in Scott's denotational semantics. In addition, we construct several other infinite CPO's. For some of those, we apply the usual Fixed Point Theorem (FPT) to yield a fixed point for every continuous function $\mu:2\rightarrow 2$ (where 2 denotes the set $\{0,1\}$), while for the other CPO's we cannot invoke that theorem to yield such fixed points.  Every element of each of these CPO's is a binary string in the monotypic form and we show that invalidation of the applicability of the FPT to the CPO that Scott's constructed yields the concept of replication. 
\end{abstract}
\end{frontmatter}
\section{Introduction}\label{SecIntro_p4}
One of the most important differences between physics and biology is the contrast between indistinguishable particles and replications.  The former is a constraint/condition on calculations in statistical mechanics, and there are two types of indistinguishable particles in the physical world:\ bosons and fermions.  Conversely, DNA replication is one example of the latter. It occurs in a cell and is followed by a cell division.  In the sense that DNA and a cell can be interpreted as a code and its decoder, respectively, there is a measurement process between them.

In his theory of denotational semantics~\cite{DScott1}, Scott regards a computer program as an element of a complete partial order (CPO).  He constructs a CPO $X$ such that $X\simeq\mathrm{C}(X,2)$, where 2 denotes the set $\{0,1\}$ and $\mathrm{C}(X,2)$ is the set of all continuous functions $f:X\rightarrow 2$. (Here, the term {\it continuous function} is used in the context of a CPO, which is defined later.) On one hand, he uses the isomorphism between $X$ and $\mathrm{C}(X,2)$ to obtain a fixed point for every continuous function $\mu:2\rightarrow 2$. (For a set~$Y$, a function $\tau:Y\rightarrow Y$ has a fixed point if there is some $y\in Y$ such that $\tau(y)=y$.) On the other hand, he regards the relation $X\simeq\mathrm{C}(X,2)$ as a domain equation, which he uses to find a new meaning for the program as a loop structure (self-similar loop program).

Gunji interprets the isomorphism as an ``abstract'' boundary to which the Fixed Point Theorem~(FPT) (Theorem~\ref{PropFixedpoint2_p4}) is applicable, and the domain equation as an invalidation of the boundary, that is, an invalidation of the applicability of the FPT, in order to construct a dynamical system in the context of theoretical biology~\cite{GunjiItoKusunoki}. He argues that any solution to a problem will inevitably be a pseudo-solution, and that the pseudo-solution at any given time step ``always'' triggers a problem to be solved at the next time step.  This leads to a perpetual evolutionary process that allows for emergent properties.  This perpetual cycle of problem and pseudo-solution is called a {\it perpetual equilibrating mechanism}~\cite{Matsuno1}, which Matsuno proposed in the field of research in theoretical biology known as Internal Measurement (IM)~\cite{Matsuno1,Matsuno2,GunjiItoKusunoki,GunjiHarunaSawa}.  Recently, Matsuno and Gunji emphasized the importance of a type of engine in theoretical biology by which dynamical systems evolve and occurrences of distinction continue to alternate with occurrences of invalidation of the distinction~\cite{Matsuno2,GunjiHarunaSawa}.

We may express an IM's model of living things in terms of a triad: two different logical layers such as the Extent (collection of fragments/elements/parts) and the Intent (property as a whole), and a mediator/interface to adjust the two layers in an ``inconsistent'' manner~\cite{Matsuno2,GunjiHarunaSawa}.  IM claims that the relationship between the two layers is not consistently determined, and for this reason they are perpetually changing relative to one another.  In one of the models in IM proposed by Gunji et al.~\cite{GunjiHarunaSawa}, first, the Extent and Intent are regarded to have the structures of a lattice and a quotient lattice, respectively.  The operations between the two layers, $\sigma:\,\mathrm{Extent}\rightarrow \mathrm{Intent}$ and $\rho:\,\mathrm{Intent}\rightarrow \mathrm{Extent}$, can be defined as a sheaf (that is, a type of integration operation) from a lattice to a quotient lattice, and a reverse sheaf (that is, a type of differentiation operation) from a quotient lattice to a lattice, respectively, only when the Extent and Intent are consistent with each other.  However, IM claims a fundamental inconsistency between the Extent and Intent.  Hence, second, an observer who cannot look out over the whole lattice (i.e., Extent) is introduced on the reverse sheaf, $\rho:\,\mathrm{Intent}\rightarrow \mathrm{Extent}$.  This reveals a collapse of the lattice structure itself in the Extent.  Third, a new mathematical operation is introduced, called a skeleton (repairing function), which is required to repair the broken structure in the Extent. By the operation of the skeleton, the lattice structure is recovered and then a new quotient lattice is constructed by the sheaf, $\sigma:\,\mathrm{Extent}\rightarrow \mathrm{Intent}$.  This process (consisting of $\rho$ and $\sigma$) is going on perpetually.  The skeleton that mediates between the two levels is regarded as a particular expression for the material cause and/or a clock as a particular expression for time itself.

One of several distinguishing features in Gunji's model (in the context of the current paper) is the role of the observer who cannot look out over the whole lattice (i.e., Extent).  On one hand, IM claims that the two layers are inconsistent with each other and are therefore changing perpetually. On the other hand, if we suppose that the defective observer can look out over the whole lattice, then the operation between the two layers can be defined consistently (as sheaf and reverse-sheaf) and the perpetual process therefore stops.  A question then arises: Are the two layers consistent or inconsistent?  If they are consistent, then the defective observer exists, unfortunately, just to destroy the structure of the Extent.  If they are inconsistent, that is, if we assume the consistency between them and we obtain a logical paradox such as Russel's paradox~\cite{Lawvere1969}, what the observer destroys is the presupposition of the consistency.  Which is correct in IM? However, it does not matter whether he is a destroyer or a life-saving super-preventer, because he cannot look out over the whole lattice in any case.  This must be the essential treatment of the logical paradox in IM.

Rosen also considers such a paradox in his model of a living thing as a {\it metabolism-repair system} ({\it M-R system})~\cite{Rosen1}.  The model consists of two sets: $X$, which is a set of raw materials, and $Y$, which is a set of {\it behaviors}. It also includes three functions: $f\in F=\mathrm{Hom}(X,$\,$Y)$, called a {\it metabolic function}, $g\in G=\mathrm{Hom}(Y,$\,$F)$, called a {\it repair function}, and $h\in H=\mathrm{Hom}(F,$\,$G)$, called a {\it replication function}, where he claims $g\in G$ and $h\in H$ are onto functions and $Y\simeq H$ holds.  One of two remarkable features in Rosen's model is that there are no such onto functions $g\in G$ and $h\in H$.  If we assume that they exist, then we obtain a paradox such as Russel's paradox~\cite{Lawvere1969}.  A second remarkable feature is that any function in the system can be both a function and an output of a different function.

He claims that a central feature of living things is {\it complexity}.  A {\it system} is called {\it complex} if its behavior cannot be captured by {\it models} of that system; otherwise, that system is called {\it simple}~\cite{Rosen3}.  The term ``complex'' for a system may be replaced with ``{\it incomputable}'' or ``{\it not well-formed}'' for its models.  Therefore a system is called complex only if its models are incomputable or not well-formed, and so on.

A treatment of the paradox is to invoke hyperset theory~\cite{Kercel2003}.  A hyperset is defined as a graphable set which is a digraph (``digraph'' is short for directed graph) such that a node can be either a set or an element of a set, and a directed edge $\rightarrow$ is the set membership $\ni$.  $A\ni A$, which leads to Russel's paradox~\cite{Lawvere1969}, is interpreted as just a loop structure in hyperset theory.  Therefore we should redefine the complexity of a system so that a system is called complex if it cannot be well-formed in Rosen's original model but can be well-formed (as a loop structure) in hyperset theory.

Let $\Pi$ be a collection of concepts/entities, and let $\mathrm{Int}(\pi)$ and $\mathrm{Ext}(\pi)$ be the intension and extension of the concept $\pi\in\Pi$. The triad---$\mathrm{Int}(\pi),\ \mathrm{Ext}(\pi)$, and $\pi\in\Pi$--- is a useful means to organize our thoughts. There are at least two usages. One is when we assume that, given a pair of intension $\mathcal{I}$ and extension $\mathcal{E}$, we try to find a concept $\pi\in\Pi$ such that $\mathcal{I}=\mathrm{Int}(\pi)$ and $\mathcal{E}=\mathrm{Ext}(\pi)$. Another is when, given a concept $\pi\in\Pi$, we try to find its $\mathrm{Int}(\pi)$ and $\mathrm{Ext}(\pi)$.

When $\mathrm{Int}(\pi)$ is given as $\mathrm{Hom}(\mathrm{Ext}(\pi),\,2)$, where $\mathrm{Hom}(A,\,B)$ denotes the set of all morphisms from $A$ to $B$, and $2$ denotes a set $\{0,\,1\}$, if we assume that $\mathrm{Int}(\pi)\simeq \mathrm{Ext}(\pi)$, then we obtain the paradox.  We can regard the paradox in an M-R system as this case. However, we have to note that there is not the associated concept $\pi\in\Pi$ in both cases of the original M-R system and its interpretation with hyperset theory. Although it is still difficult to specify what the associated $\pi\in\Pi$ in IM is, we may at least state that it is material causatic. 

In this paper we will be consistent and consider binary strings in the form of:
\begin{equation}\label{Eq_FormBinaryString_p4}
\underbrace{0\cdots 0}_{u}\underbrace{1\cdots 1}_{v},
\end{equation}
where $u$ is the number of 0's and $v$ is the number of 1's, and $u,v\in \mathcal{N}=\{0,1,2,\ldots\}$. We will regard (\ref{Eq_FormBinaryString_p4}) as the concept $\pi\in\Pi$ that we will consistently consider in this paper. When we insert a comma at the center of a binary string to divide the string into two parts, we regard the left part as the $\mathrm{Int}(\pi)$ and the right part as the $\mathrm{Ext}(\pi)$.  We should note that the notation ``$0\cdots 01\cdots 1$'' means that the string consists of finitely many 0's, followed by finitely many 1's.  Hence, we must explicitly define binary strings one-by-one, such as
\begin{itemize}
\item finitely many 0's, followed by finitely many 1's
\item infinitely many 0's, followed by finitely many 1's
\item finitely many 0's, followed by infinitely many 1's
\item[]$\cdots$ etc.
\end{itemize}
We designate these strings by (\ref{Eq_FormBinaryString_p4}) in this paper.  In this sense, both $\pi\in\Pi$ and (\ref{Eq_FormBinaryString_p4}) are not strict mathematical statements.

We will construct various infinite CPO's.  In each of the CPO's an element (i.e., a binary string) is in the form of (\ref{Eq_FormBinaryString_p4}).  For some of those CPO's, we can apply the FPT to yield a fixed point for every continuous function $\mu:2\rightarrow 2$, while for the other CPO's we cannot invoke that theorem to yield such fixed points.  In section~\ref{SecConstFixedPoint_p4}, we construct a countably infinite CPO $S$ in the manner of Scott's denotational semantics, and we define an isomorphism from $S$ to $\mathrm{C}(S,2)$ in the usual way.  In section~\ref{SecAnoterRepresentation_p4}, we construct a CPO which is isomorphic to $S$, by using a completely different method. Then in section~\ref{SecLR_p4}, we introduce a transformation (called the LR-transformation) between certain types of infinite binary strings that are defined from different specifications of certain finite binary strings.  In section~\ref{SecRep_p4} we show that the concept of replication is derived from invalidating the applicability of the FPT to the CPO that Scott constructed.

\section{Construction of infinite CPO}\label{SecConstFixedPoint_p4}
\subsection{Preliminaries}\label{SubSecPreliminary_p4}

In this paper, we construct a number of linear orders, each of which is a complete partial order.
\begin{defn}\label{DefCPO_p4}     A partial order $(D,\subseteq)$ is a \emph{complete partial order }(CPO) if $D$ has a $\subseteq$-least element and every countable, monotone non-decreasing sequence $ d_{0}\subseteq d_{1}\subseteq d_{2}\subseteq\cdots$ of elements of $D$ has a unique least upper bound $\cup d_{i}$ in $D$.
\end{defn}

\begin{defn}\label{DefContinuity_p4} Let $(D,\subseteq_{D})$ and $(E,\subseteq_{E})$ be CPO's.  A function $g:D\rightarrow E$ is \emph{continuous}         if it satisfies the following conditions:\\
(1) If $d_{i}\subseteq_{D}d_{j}$, then $g(d_{i})\subseteq_{E}g(d_{j})$.\\
(2) If $ d_{0}\subseteq d_{1}\subseteq d_{2}\subseteq\cdots$ is a countable, monotone non-decreasing sequence of elements of $D$, then $g(\cup d_{i})=\cup(g(d_{i}))$.
\end{defn}
\begin{defn}\label{DefEmbeddingProjection_p4}             
The relation $D\subseteq E$ holds of CPO's $D,\, E$ if there exist continuous functions $e:D\rightarrow E$ and $p:E\rightarrow D$ such that $e\circ p\subseteq\mathrm{id}_{E}$ and $p\circ e=\mathrm{id}_{D}$. The functions $e$ and $p$ are called \emph{embedding}   and \emph{projection}, respectively.  If $e\circ p=\mathrm{id}_{E}$, then $D$ is \emph{isomorphic}   to $E$, which is denoted by $D\simeq E$.
\end{defn}

\begin{prop}\cite{DScott1}\label{PropDScott1_p4}
Let $D$ and $E$ be CPO's, and let $\mathrm{C}(D,E)$ be the set of all continuous functions from $D$ to $E$.  Then $(\mathrm{C}(D,E),\subseteq)$ is also a CPO, the least element being the constant function which maps every $d\in D$ to the least element of $E$, and the least upper bound of the countable, monotone non-decreasing sequence $ g_{0}\subseteq g_{1}\subseteq g_{2}\subseteq\cdots$ being the function $\cup g_{i}$ defined by $(\cup g_{i})(d)=\cup(g_{i}(d))$.
\end{prop}

Let 2 be the set $\{0,1\}$. Clearly, the linear order $(2,\subseteq)$ defined by $0\subseteq 1$ is a CPO (which we will denote by 2), so the next result follows immediately from Proposition~\ref{PropDScott1_p4}.

\begin{cor}\label{PropCPOs1_p4}
Let $(S_{0},\subseteq)$ be a CPO. Then $S_{1}\equiv\mathrm{C}(S_{0},2),$\ $ S_{2}\equiv\mathrm{C}(\mathrm{C}(S_{0},2),2),\ldots$ are also CPO's.
\end{cor}

\begin{thm}\label{PropFixedpoint2_p4}(Fixed Point Theorem)~\cite{Hasegawa,Gunter}             
Let $\mu:2\rightarrow 2$ be continuous, let $(S,\subseteq)$ be a  CPO which is isomorphic to $\mathrm{C}(S,2)$, and let $\varphi$ be an isomorphism (a continuous bijection) from $S$ to $\mathrm{C}(S,2)$. Furthermore, let $g:S\rightarrow 2$ be the continuous function defined by\[g(x)=\mu(\hat{\varphi}_{x}(x)),\] where $\hat{\varphi}_{x}$ denotes $\varphi(x)$. Then $g(\varphi^{-1}(g))$ is a fixed point of $\mu$.
\end{thm}

{\bf Proof}\,\,\, By the definition of $\varphi$, we have that, for every $f\in\mathrm{C}(S,2)$ and every $x\in S$,\[f(x)=\hat{\varphi}_{\varphi^{-1}(f)}(x)\]
Since $g\in\mathrm{C}(S,2)$, we have that, for every $x\in S$,\[g(x)=\hat{\varphi}_{\varphi^{-1}(g)}(x)\]
Setting $x$ to $\varphi^{-1}(g)$ yields\[g(\varphi^{-1}(g))=\hat{\varphi}_{\varphi^{-1}(g)}(\varphi^{-1}(g))\]
Also, by the definition of $g$, we have \[g(\varphi^{-1}(g))=\mu(\hat{\varphi}_{\varphi^{-1}(g)}(\varphi^{-1}(g)))\]
 Thus\[\mu(\hat{\varphi}_{\varphi^{-1}(g)}(\varphi^{-1}(g)))=g(\varphi^{-1}(g))=\hat{\varphi}_{\varphi^{-1}(g)}(\varphi^{-1}(g))\]
Hence $\hat{\varphi}_{\varphi^{-1}(g)}(\varphi^{-1}(g))$ ($=g(\varphi^{-1}(g))$) is a fixed point of $\mu$.  $\blacksquare$

Preparatory to constructing an infinite CPO, we construct an infinite sequence of finite CPO's. We do this in stages, which we index with the natural numbers $n\ge 1$. (Throughout this paper, the variable $n$ denotes a natural number, i.e., an element of the set $\mathcal{N}=\{0,1,2,\ldots\}$. Except where $n$ is specifically restricted in some way---such as in the stages of this construction, where $n$ is restricted to values greater than 0---it is to be assumed that $n$ is any nonnegative integer.)

At stage $n$, we construct a set $S_{n}$ with $n$ elements, and we define a linear order $\subseteq_{n}$ on $S_{n}$. Since $S_{n}$ is finite, $(S_{n},\subseteq_{n})$ is a CPO.  We represent the elements of $S_{n}$ as binary strings (strings of 0's and 1's) of length $n-1$. Later, we will use infinite sequences of elements of $\bigcup_{n=1}^\infty S_{n}$ to construct two infinite sets $S$. In each case, we will define a linear order $\subseteq$ on $S$ in such a way that $(S,\subseteq)$ is a CPO.

\begin{enumerate}
\item[\textit{Stage 1}]Let $S_{1}=\{\lambda\}$, where $\lambda$ is the empty string (the binary string of length 0). Then $(S_{1},\subseteq_{1})$ is trivially a CPO (where $\lambda\subseteq_{1}\lambda$). We will denote $(S_{1},\subseteq_{1})$ by $\{\lambda\}$.
\item[\textit{Stage 2}]Let $S_{2}=\mathrm{C}(S_{1},2)$, and let $(S_{2},\subseteq_{2})$ be the CPO obtained from Proposition~\ref{PropDScott1_p4}       for $D=\mathrm{C}(S_{1},2)$ and $E=2$. Since $S_{2}$ is the set of continuous functions from $\{\lambda\}$ to $2$ (where $2=\{0,1\}$), $S_{2}$ consists of the functions $(\lambda\rightarrow 0)$ and $(\lambda\rightarrow 1)$.  We will denote those functions by their outputs ($0$ and $1$, respectively).  Recall that $0\subseteq 1$ (by our definition of the CPO 2), so we will denote $(S_{2},\subseteq_{2})$ by $\{0\subseteq 1\}$.
\item[\textit{Stage 3}]Similarly, let $S_{3}=\mathrm{C}(S_{2},2)$, and let $(S_{3},\subseteq_{3})$ be the CPO obtained from Proposition~\ref{PropDScott1_p4}       for $D=\mathrm{C}(S_{2},2)$ and $E=2$. Since $S_{3}$ is the set of continuous functions from $S_{2}$ to $2,\, S_{3}$ consists of the functions $(0\rightarrow 0;1\rightarrow 0),\,(0\rightarrow 0;1\rightarrow 1)$, and $(0\rightarrow 1;1\rightarrow 1)$.  The function $(0\rightarrow 1$; $1\rightarrow 0)$ is excluded, because it is not continuous (the outputs do not preserve the order of the inputs).  We will denote the three elements of $S_{3}$ by $00,\, 01$, and $11$, respectively (i.e., for each element of $S_{3}$---equivalently, for each continuous function from $S_{2}$ into 2---we concatenate the outputs that correspond to the two inputs, 0 and 1), so we will denote $(S_{3},\subseteq_{3})$ by $\{00\subseteq 01\subseteq 11\}$.
\item[$\vdots$]{}
\item[\textit{Stage $n$}]Let $S_{n}=\mathrm{C}(S_{n-1},2)$, and let $(S_{n},\subseteq_{n})$ be the CPO obtained from Proposition~\ref{PropDScott1_p4}   for $D=\mathrm{C}(S_{n-1},2)$ and $E=2$.
\end{enumerate}

\begin{eqnarray}\label{EqConstructionHigherOrderedSet_p4}
\begin{array}{ccl}
(S_{1},\subseteq_{1})&=&\{\lambda\}\\
(S_{2},\subseteq_{2})&=&\{0\subseteq_{2}1\}\\
(S_{3},\subseteq_{3})&=&\{00\subseteq_{3}01\subseteq_{3}11\}\\
(S_{4},\subseteq_{4})&=&\{000\subseteq_{4}001\subseteq_{4}011\subseteq_{4}111\}\\
&\vdots&\\
(S_{n},\subseteq_{n})&=&\{\underbrace{0\cdots 0000}_{n-1}\subseteq_{n}\underbrace{0\cdots 0001}_{n-1}\subseteq_{n}\underbrace{0\cdots 0011}_{n-1}\subseteq_{n}\underbrace{0\cdots 0111}_{n-1}\subseteq_{n}\cdots\\
&&\subseteq_{n}\underbrace{0001\cdots 1}_{n-1}\subseteq_{n}\underbrace{0011\cdots 1}_{n-1}\subseteq_{n}\underbrace{0111\cdots 1}_{n-1}\subseteq_{n}\underbrace{1111\cdots 1}_{n-1}
\end{array}
\end{eqnarray}

For fixed $n$, there may be several ways to define the embedding and projection functions $e:S_{n}\rightarrow S_{n+1}$ and $p:S_{n+1}\rightarrow S_{n}$, and different embedding and projection functions will lead to different infinite sets $S$.  We now review the standard definitions of embedding and projection, as used by Scott~\cite{DScott1} and others~\cite{SotoAndradeVarela1984,Hasegawa,Gunter}.

\subsection{Standard definitions of embedding and projection}\label{SubSecUsualWayOfConstruction_p4}
Let $ S_{1},S_{2},S_{3},\ldots$ be the finite sets defined in~(\ref{EqConstructionHigherOrderedSet_p4}). We now present two different definitions of the embedding and projection functions. For each definition, we use the projection functions to generate a countably infinite CPO.

For every $n\ge 1$, define the embedding function $e_{n}:S_{n}\rightarrow S_{n+1}$ and the projection function $p_{n}:S_{n+1}\rightarrow S_{n}$ as shown in Table~\ref{TableEmbeddingProjection_p4}. (Note that these functions satisfy Definition~\ref{DefEmbeddingProjection_p4}.) 

\begin{table}[htbp]\scriptsize\begin{center}
\begin{tabular}{ccccccccccc}
&&&&&&&&&&11111\\
&&&&&&&&&$\swarrow\nearrow$&\\
&&&&&&&&1111&&01111\\
&&&&&&&$\swarrow\nearrow$&&$\swarrow\nearrow$&\\
&&&&&&111&&0111&&00111\\
&&&&&$\swarrow\nearrow$&&$\swarrow\nearrow$&&$\swarrow$&\\
&&&&11&&011&&0011&$\leftrightarrow$&00011\\
&&&$\swarrow\nearrow$&&$\swarrow$&&$\swarrow$&&&\\
&&1&&01&$\leftrightarrow$&001&$\leftrightarrow$&0001&$\leftrightarrow$&00001\\
&$\swarrow$&&$\swarrow$&&&&&&&\\
$\lambda$&$\leftrightarrow$&0&$\leftrightarrow$&00&$\leftrightarrow$&000&$\leftrightarrow$&0000&$\leftrightarrow$&00000\\
$S_{1}$&&$S_{2}$&&$S_{3}$&&$S_{4}$&&$S_{5}$&&$S_{6}$\\
\end{tabular}\end{center}
\caption{Definition of embedding and projection for $n=1,2,\ldots,5$.}
\label{TableEmbeddingProjection_p4}
\end{table}

For example, $e_{2}:S_{2}\rightarrow S_{3}$ is defined as $e_{2}(0)=00,\, e_{2}(1)=11$, and $p_{2}:S_{3}\rightarrow S_{2}$ is defined as $p_{2}(00)=p_{2}(01)=0,\, p_{2}(11)=1$.

Now label all the entries in Table~\ref{TableEmbeddingProjection_p4}: Label entry $s$ with the number of 1's in $s$  (i.e., label $\lambda$,  0, 00, 000, $\ldots$ with 0; label 1, 01, 001, 0001, $\ldots$ with 1; label 11, 011, 0011, 00011, $\ldots$ with 2; etc.).  This yields Table~\ref{TableEmbeddingProjection2_p4}. 

\begin{table}[htbp]\scriptsize\begin{center}
\begin{tabular}{ccccccccccc}            
&&&&&&&&&&5\\
&&&&&&&&&$\swarrow\nearrow$&\\
&&&&&&&&4&&4\\
&&&&&&&$\swarrow\nearrow$&&$\swarrow\nearrow$&\\
&&&&&&3&&3&&3\\
&&&&&$\swarrow\nearrow$&&$\swarrow\nearrow$&&$\swarrow$&\\
&&&&2&&2&&2&$\leftrightarrow$&2\\
&&&$\swarrow\nearrow$&&$\swarrow$&&$\swarrow$&&&\\
&&1&&1&$\leftrightarrow$&1&$\leftrightarrow$&1&$\leftrightarrow$&1\\
&$\swarrow$&&$\swarrow$&&&&&&&\\
0&$\leftrightarrow$&0&$\leftrightarrow$&0&$\leftrightarrow$&0&$\leftrightarrow$&0&$\leftrightarrow$&0\\
$S_{1}$&&$S_{2}$&&$S_{3}$&&$S_{4}$&&$S_{5}$&&$S_{6}$\\
\end{tabular}\end{center}
\caption{Table~\ref{TableEmbeddingProjection_p4} after labeling of the entries}.
\label{TableEmbeddingProjection2_p4}
\end{table}

Our countably infinite CPO will consist of all the infinite paths in Table~\ref{TableEmbeddingProjection2_p4} that have the 0 in $S_{1}$ as their inverse (projective) limit (i.e., every element of our CPO is an infinite sequence $(s_{1},s_{2},s_{3},\ldots)$ such that, for every $i\ge 1,\,\, s_{i}\in S_{i}$ and $s_{i}=p_{i}(s_{i+1})$). Label every such infinite path, either with $n$ or $n^{\prime}$ for some $n\in\mathcal{N}$, or with $\infty$, as follows:

\begin{equation}\label{EqScottAssignment_p4}
\begin{array}{ccc}
0,0,0,0,0,0,0,0,0,0,\ldots&\leftrightarrow& 0\\
0,0,1,1,1,1,1,1,1,1,\ldots&\leftrightarrow& 1\\
0,0,1,1,2,2,2,2,2,2,\ldots&\leftrightarrow& 2\\
\vdots&&\\
\underbrace{0,0}_2,\underbrace{1,1}_2,\underbrace{2,2}_2,\underbrace{3,3}_2,\underbrace{4,4}_2,\ldots&\leftrightarrow&\infty\\
\vdots&&\\
0,0,1,1,2,3,4,5,6,7,\ldots&\leftrightarrow& 2^{\prime}\\
0,0,1,2,3,4,5,6,7,8,\ldots&\leftrightarrow& 1^{\prime}\\
0,1,2,3,4,5,6,7,8,9,\ldots&\leftrightarrow& 0^{\prime}\\
\end{array}
\end{equation}

The infinite path which is labeled with $\infty$ is the least upper bound of the infinite paths labeled with $n$ for some $n\in\mathcal{N}$. Thus we have a countably infinite CPO with the following structure:
\begin{equation}\label{EqVinf_p4}
\Phi=\{0\subseteq 1\subseteq 2\subseteq 3\subseteq\cdots\subseteq\infty\subseteq\cdots\subseteq 3^{\prime}\subseteq 2^{\prime}\subseteq 1^{\prime}\subseteq 0^{\prime}\}
\end{equation}

For every $n$, map the elements of $\Phi$ which are labeled with $n$ and $n^{\prime}$ to the continuous functions $\psi_{n}$ and $\psi_{n^{\prime}}$, respectively (and map the element of $\Phi$ which is labeled with $\infty$ to the continuous function $\psi_{\infty}$), as shown in Table~\ref{TableFixedpoint_p4}.
\begin{table}[htbp]
\begin{center}\scriptsize
\begin{tabular}{|c|ccccccccccccc|} \hline
$\psi\backslash x$ & 0 & 1 & $\cdots$ & $n-1$ & $n$ & $\cdots$& $\infty$&$\cdots$ & $(n-1)^{\prime}$ & $n^{\prime}$ & $\cdots$  & $1^{\prime}$ & $0^{\prime}$ \\ \hline
$\psi_{0}$ & 0 & 0 & $\cdots$ & 0 & 0 & $\cdots$ & 0&$\cdots$ & 0 & 0 & $\cdots$ & 0 & 0 \\
$\psi_{1}$ & 0 & 0 & $\cdots$ & 0 & 0 & $\cdots$ & 0&$\cdots$ & 0 & 0 & $\cdots$ & 0 & 1 \\
$\vdots$ &  &  &  &  &  &  &  &  &  &  &  &  & \\
$\psi_{n}$ & 0 & 0 & $\cdots$ & 0 & 0 & $\cdots$ & 0&$\cdots$ & 0 & 1 & $\cdots$ & 1 & 1 \\
$\vdots$ &  &  &  &  &  &  &  &  &  &  &  &  & \\
$\psi_{\infty}$ & 0 & 0 & $\cdots$ & 0 & 0 & $\cdots$ & 0&$\cdots$ & 1 & 1 & $\cdots$ & 1 & 1\\ 
$\vdots$ &  &  &  &  &  &  &  &  &  &  &  &  & \\
$\psi_{n^{\prime}}$ & 0 & 0 & $\cdots$ & 0 & 1 & $\cdots$ & 1&$\cdots$ & 1 & 1 & $\cdots$ & 1 & 1 \\
$\vdots$ &  &  &  &  &  &  &  &  &  &  &  &  & \\
$\psi_{1^{\prime}}$ & 0 & 1 & $\cdots$ & 1 & 1 & $\cdots$ & 1&$\cdots$ & 1 & 1 & $\cdots$ & 1 & 1 \\
$\psi_{0^{\prime}}$ & 1 & 1 & $\cdots$ & 1 & 1 & $\cdots$ & 1&$\cdots$ & 1 & 1 & $\cdots$ & 1 & 1 \\ \hline
\end{tabular}
\end{center}
\caption{The functions $\psi$ in $\mathrm{C}(\Phi,2)$ and their values ($\psi(x)$ for $x\in\Phi$)}
\label{TableFixedpoint_p4}
\end{table}

From the table, it is obvious that the map $\hat{\psi}:\Phi\rightarrow\mathrm{C}(\Phi,2)$ which consists of the union of the maps $n\mapsto\psi_{n},\ \ n^{\prime}\mapsto\psi_{n^{\prime}}$, and $\infty\mapsto\psi_{\infty}$ preserves the ordering of $\Phi$. Hence $\hat{\psi}$ is a continuous function and an isomorphism.

This is the usual construction of a countably infinite CPO $\Phi$ (i.e., the construction used by Scott). Moreover, the usual derivation of a fixed point for a continuous function $\mu:2\rightarrow 2$ is that which is given in the statement of Theorem~\ref{PropFixedpoint2_p4}. We obtain that fixed point (namely, $g(\hat{\psi}^{-1}(g))$, where $g$ is as in the statement of Theorem~\ref{PropFixedpoint2_p4}) from Table~\ref{TableFixedpoint_p4}: 

\begin{itemize}
\item if $\mu(0)=0=\mu(1)$, then $g$ is $\psi_{0}$, so the fixed point is $\psi_{0}(0)$ ($=0$)
\item if $\mu(0)=1=\mu(1)$, then $g$ is $\psi_{0^{\,\prime}}$, so the fixed point is $\psi_{0^{\,\prime}}(0^{\prime})$ ($=1$)
\item if $\mu(0)=0$ and $\mu(1)=1$, then $g$ is $\psi_{\infty}$, so the fixed point is $\psi_{\infty}(\infty)$ ($=0$)
\end{itemize}

As an alternative, we can define the embedding and projection functions as in Table~\ref{TableEmbeddingProjection3_p4}, and we can use those functions to define a countably infinite CPO which has a structure different from that of $\Phi$.

\begin{table}[htbp]\scriptsize\begin{center}
\begin{tabular}{ccccccccccc}            
&&&&&&&&&&11111\\
&&&&&&&&&$\swarrow$&\\
&&&&&&&&1111&$\leftrightarrow$&01111\\
&&&&&&&$\swarrow$&&&\\
&&&&&&111&$\leftrightarrow$&0111&$\leftrightarrow$&00111\\
&&&&&$\swarrow$&&&&&\\
&&&&11&$\leftrightarrow$&011&$\leftrightarrow$&0011&$\leftrightarrow$&00011\\
&&&$\swarrow$&&&&&&&\\
&&1&$\leftrightarrow$&01&$\leftrightarrow$&001&$\leftrightarrow$&0001&$\leftrightarrow$&00001\\
&$\swarrow$&&&&&&&&&\\
$\lambda$&$\leftrightarrow$&0&$\leftrightarrow$&00&$\leftrightarrow$&000&$\leftrightarrow$&0000&$\leftrightarrow$&00000\\
$S_{1}$&&$S_{2}$&&$S_{3}$&&$S_{4}$&&$S_{5}$&&$S_{6}$\\
\end{tabular}\end{center}
\caption{Alternative definition of embedding and projection for $n=1,2,\ldots,5$}.
\label{TableEmbeddingProjection3_p4}
\end{table}

Using the same labeling scheme in Table~\ref{TableEmbeddingProjection3_p4}    as that which was used in Table~\ref{TableEmbeddingProjection_p4} yields Table~\ref{TableEmbeddingProjection4_p4}.

\begin{table}[htbp]\scriptsize\begin{center}
\begin{tabular}{ccccccccccc}            
&&&&&&&&&&5\\
&&&&&&&&&$\swarrow$&\\
&&&&&&&&4&$\leftrightarrow$&4\\
&&&&&&&$\swarrow$&&&\\
&&&&&&3&$\leftrightarrow$&3&$\leftrightarrow$&3\\
&&&&&$\swarrow$&&&&&\\
&&&&2&$\leftrightarrow$&2&$\leftrightarrow$&2&$\leftrightarrow$&2\\
&&&$\swarrow$&&&&&&&\\
&&1&$\leftrightarrow$&1&$\leftrightarrow$&1&$\leftrightarrow$&1&$\leftrightarrow$&1\\
&$\swarrow$&&&&&&&&&\\
0&$\leftrightarrow$&0&$\leftrightarrow$&0&$\leftrightarrow$&0&$\leftrightarrow$&0&$\leftrightarrow$&0\\
$S_{1}$&&$S_{2}$&&$S_{3}$&&$S_{4}$&&$S_{5}$&&$S_{6}$\\
\end{tabular}\end{center}
\caption{Table~\ref{TableEmbeddingProjection3_p4} after labeling of the entries}
\label{TableEmbeddingProjection4_p4}
\end{table}

Now label every infinite path in Table~\ref{TableEmbeddingProjection4_p4} that has the 0 in $S_{1}$ as its inverse (projective) limit, either with some $n\in\mathcal{N}$ or with $\infty$, as follows:
\begin{equation}
\nonumber
\begin{array}{ccc}
0,0,0,0,0,0,0,0,0,0,\ldots&\leftrightarrow& 0\\
0,1,1,1,1,1,1,1,1,1,\ldots&\leftrightarrow& 1\\
0,1,2,2,2,2,2,2,2,2,\ldots&\leftrightarrow& 2\\
0,1,2,3,3,3,3,3,3,3,\ldots&\leftrightarrow& 3\\
\vdots&&\\
\underbrace{0,1,2,\cdots,n-1}_{n},n,n,\ldots&\leftrightarrow& n\\
\vdots&&\\
0,1,2,3,4,5,6,7,8,9,\ldots&\leftrightarrow&\infty\\
\end{array}
\end{equation}

This yields a CPO with the following structure:
\begin{equation}\label{EqVinf3_p4}
\Theta=\{0\subseteq 1\subseteq 2\subseteq 3\subseteq\cdots\subseteq\infty\}
\end{equation}

Unfortunately, $\Theta$ is not isomorphic to $\mathrm{C}(\Theta,2)$. To see this, we will examine the only two mappings from $\Theta$ to $\mathrm{C}(\Theta,2)$ that could possibly be isomorphisms.

One possibility is that, for every $n\in\mathcal{N}$, the element of $\Theta$ which is labeled by $n$ would be mapped to the function $\psi_{n}^{\prime}\in\mathrm{C}(S,2)$ defined by $\psi_{n}^{\prime}(x)=1$ if $x$ is one of the $n$ greatest elements of $\Theta$, and $\psi_{n}^{\prime}(x)=0$ otherwise (and mapping the element of $\Theta$ which is labeled by $\infty$ to the all-1 function $\psi_{\infty}^{\prime}$). However, the ``$n$ greatest elements of $\Theta$'' do not exist, so this mapping is impossible.

Alternatively, we could map the element of  $\Theta$ which is labeled by $n$ to the function $\psi_{n}^{\prime\prime}\in\mathrm{C}(\Theta,2)$ defined by $\psi_{n}^{\prime\prime}(x)=1$ for every $ x\in\Theta$ with $n\subseteq x$, and $\psi_{n}^{\prime\prime}(x)=0$ otherwise (and mapping the element which is labeled by $\infty$ to the all-0 function $\psi_{\infty}^{\prime\prime}$), as shown in Table~\ref{TableFixedpoint3_p4}. Note, however, that $\psi_{y}^{\prime\prime}\subseteq\psi_{x}^{\prime\prime}$ for elements $x,y$ of $\Theta$ with $x\subseteq y$, so this mapping does not preserve the ordering of $\Theta$; hence it is not an isomorphism.

\begin{table}[htbp]
\begin{center}\scriptsize
\begin{tabular}{|c|cccc|} \hline
$\psi^{\prime\prime}\backslash x$ & 0 & 1 & 2 & $\cdots$ \\ \hline
$\psi_{0}^{\prime\prime}$ & 1 & 1 & 1 & $\cdots$ \\
$\psi_{1}^{\prime\prime}$ & 0 & 1 & 1 & $\cdots$ \\
$\psi_{2}^{\prime\prime}$ & 0 & 0 & 1 & $\cdots$ \\
$\vdots$ &  &  &  & \\
$\psi_{\infty}^{\prime\prime}$ & 0 & 0 & 0 & $\cdots$ \\ \hline
\end{tabular}
\end{center}
\caption{The functions $\psi^{\prime\prime}$ in $\mathrm{C}(\Theta,2)$ and their values ($\psi^{\prime\prime}(x)$ for $x\in\Theta$)}
\label{TableFixedpoint3_p4}
\end{table}

\section{Another representation of $\Phi$ and $\Theta$, and beyond}\label{SecAnoterRepresentation_p4}
We now use a different construction for an infinite CPO that has the same structure as $\Phi$, and then we go on to construct infinite CPO's whose structures differ from that of $\Phi$.
\subsection{Non-standard representations of $\Phi$ and $\Theta$}\label{SubSecNonStandardRep_p4}
First, we construct a countably infinite CPO $\Lambda$ which is isomorphic to $\Phi$. Each element of $\Lambda$ is an infinite binary string. We do this directly, without first constructing a sequence of finite CPO's.

Denote the order type of the usual linear order on $\mathcal{N}$ (i.e., $ 0\subseteq 1\subseteq 2\subseteq\cdots$) by $\omega$, and denote the reverse order (i.e., $\cdots\subseteq 2\subseteq 1\subseteq 0$) by $\omega^{*}$.

To construct $\Lambda$, we first obtain one infinite set of infinite binary strings, $\Omega$, by starting with the all-0 string that has order type $\omega^{*}$ (i.e., the all-0 string $\cdots 000$), and generating the remaining strings in succession by changing the rightmost 0 to a 1:\[\Omega=\{\cdots 000,\ \cdots 001,\ \cdots 0011,\ \cdots 00111,\ \ldots\}\]

This gives us a linear order that has order type $\omega$:  

\begin{equation}\label{EqAscendingSequence_p4}
\Omega=\{\ \cdots 000\,\subseteq\,\cdots 001\,\subseteq\,\cdots 0011\,\subseteq\,\cdots 00111\,\subseteq\ \cdots\ \}
\end{equation}

Next, we obtain another infinite set of binary strings, $\Omega^{\mathrm{opp}}$, by starting with the all-1 string that has order type $\omega$ (i.e., the all-1 string $ 111\cdots$), and generating the remaining strings in succession by changing the leftmost 1 to a 0:\[\Omega^{\mathrm{opp}}=\{111\cdots\!,\ 011\cdots\!,\ 0011\cdots\!,\ 00011\cdots\!,\ \ldots\}\]

This gives us a linear order that has order type $\omega^{*}$:  
\begin{equation}\label{EqDescendingSequence_p4}
\Omega^{\mathrm{opp}}=\{\ \cdots\ \subseteq\, 00011\cdots\,\subseteq\, 0011\cdots\,\subseteq\, 011\cdots\,\subseteq 111\cdots\ \}
\end{equation}

Label $ x\in\Omega$ with $n$, where $n$ is the number of 1's in $x$. Similarly, label $y\in\Omega^{\mathrm{opp}}$ with $n^{\prime}$, where $n$ is the number of 0's in $y$. Then we have a linear order $(\Omega\cup\Omega^{\mathrm{opp}},\subseteq)$ of order type $\omega+\omega^{*}$:\[0\subseteq 1\subseteq 2\subseteq\cdots\subseteq n\subseteq\cdots\subseteq\cdots\subseteq n^{\prime}\subseteq\cdots\subseteq 2^{\prime}\subseteq 1^{\prime}\subseteq 0^{\prime}\]

This is not a CPO, because $\Omega$ has no ``sup'' (supremum, i.e., least upper bound). The most natural infinite binary string we could use as the sup of $\Omega$ is the all-1 string that has order type $\omega^{*}$ (i.e., the string $\cdots 111$).  If we place that all-1 string between the elements of $\Omega$ and the elements of $\Omega^{\mathrm{opp}}$ (and label it with $\infty$), we obtain the following CPO:

\begin{eqnarray}\nonumber
\Lambda&=&\Omega^{\prime}\,\cup\,\Omega^{\mathrm{opp}}
\label{EqVinf2_p4}\\
[.05in]&=&\{0\subseteq 1\subseteq 2\subseteq 3\subseteq\cdots\subseteq\infty\subseteq\cdots\subseteq 3^{\prime}\subseteq 2^{\prime}\subseteq 1^{\prime}\subseteq 0^{\prime}\},
\end{eqnarray}
where $\Omega^{\prime}=\Omega\cup\{\cdots 111\}$. This CPO has order type $\omega+1+\omega^{*}$, as does $\Phi$ in~(\ref{EqVinf_p4}), so $\Lambda\simeq\Phi$.

We can assign to $ x\in\Lambda$ the function $\psi_{x}$ in Table~\ref{TableFixedpoint_p4}.  Hence the map $x\mapsto\psi_{x}$ is a continuous function and an isomorphism from $\Lambda$ to $\mathrm{C}(\Lambda,2)$.  For every continuous function $\mu:2\rightarrow 2$, we obtain the same fixed point as the one we found earlier (where we applied Theorem~\ref{PropFixedpoint2_p4} to the isomorphism $\hat{\psi}:\Phi\rightarrow\mathrm{C}(\Phi,2)$). 

\begin{defn}
Let $x$ be an element of a linear order $(S,\subseteq)$. Then $x$ has an immediate neighbor in $(S,\subseteq)$ if $x$ has an immediate predecessor and/or an immediate successor in $(S,\subseteq)$. For example, both $\cdots 000$ and $ 111\cdots$ have an immediate neighbor in $\Lambda$ ($\cdots 000$ has an immediate successor (namely, $\cdots 001$), and $ 111\cdots$ has an immediate predecessor (namely, $ 011\cdots$)), but $\cdots 111$ has no immediate neighbor in $\Lambda$.
\end{defn}

Clearly, $\Omega^{\prime}\simeq\Theta$. However, $\Omega^{\prime}\not\simeq\Omega^{\mathrm{opp}}$, because  $\Omega^{\prime}$ has order type $\omega+1$ and $\Omega^{\mathrm{opp}}$ has order type $\omega^{*}$. Since \[\Omega^{\prime}\simeq\Theta\not\simeq\mathrm{C}(\Theta,2)\simeq\mathrm{C}(\Omega^{\prime},2),\] we have $\Omega^{\prime}\not\simeq\mathrm{C}(\Omega^{\prime},2)$.  Thus, for a continuous function $\mu:2\rightarrow 2$, we cannot apply Theorem~\ref{PropFixedpoint2_p4} to yield a fixed point of $\mu$.

\subsection{CPO of order type $\omega+1+1+\omega^{*}$; adjunction}\label{SubSecOderType2_p4}
In $\Lambda$, the infinite binary string $\cdots 111$ is not only the sup of $\Omega$ but also the ``inf'' (infimum, i.e., greatest lower bound) of $\Omega^{\mathrm{opp}}$. However, the most natural infinite binary string we could use as the inf of $\Omega^{\mathrm{opp}}$ is the all-0 string that has order type $\omega$  (i.e., the string $ 000\cdots$).  If we place that all-0 string between the elements of $\Omega^{\prime}$ and the elements of $\Omega^{\mathrm{opp}}$ (and label it with $\infty^{\prime}$), we obtain the following CPO:
\begin{eqnarray}\nonumber
\Lambda^{\prime}&=&\Omega^{\prime}\cup\Omega^{\prime(\mathrm{opp})}\\
[.05in]&=&\{0\subseteq 1\subseteq 2\subseteq 3\subseteq\cdots\subseteq\infty\subseteq\infty^{\prime}\subseteq\cdots\subseteq 3^{\prime}\subseteq 2^{\prime}\subseteq 1^{\prime}\subseteq 0^{\prime},\}
\end{eqnarray}
where $\Omega^{\prime(\mathrm{opp})}=\{000\cdots\}\cup\Omega^{\mathrm{opp}}$. This CPO has order type $\omega+1+1+\omega^{*}$, so it is not isomorphic to any of the linear orders discussed earlier in this paper.

We can assign to $x\in\Lambda^{\prime}$ the continuous function $\psi_{x}$ in Table~\ref{TableFixedpoint2_p4}.  Hence the map $x\mapsto\psi_{x}$ is a continuous function and an isomorphism from $\Lambda^{\prime}$ to $\mathrm{C}(\Lambda^{\prime},2)$.  For every continuous function $\mu:2\rightarrow 2$, we obtain a fixed point (namely, $g(\hat{\psi}^{-1}(g))$, where $g$ is as in the statement of Theorem~\ref{PropFixedpoint2_p4}) from Table~\ref{TableFixedpoint2_p4}: 

\begin{itemize}
\item if $\mu(0)=0=\mu(1)$, then $g$ is $\psi_{0}$, so the fixed point is $\psi_{0}(0)$ ($=0$)
\item if $\mu(0)=1=\mu(1)$, then $g$ is $\psi_{0^{\,\prime}}$, so the fixed point is $\psi_{0^{\,\prime}}(0^{\prime})$ ($=1$)
\item if $\mu(0)=0$ and $\mu(1)=1$, then $g$ is $\psi_{\infty^{\prime}}$, so the fixed point is $\psi_{\infty^{\prime}}(\infty^{\prime})$ ($=1$)
\end{itemize}

\begin{table}[htbp]\begin{center}\scriptsize
\begin{tabular}{|c|cccccccccccccc|}\hline
$\psi\backslash x$ & 0 & 1 & $\cdots$ & $n-1$ & $n$ & $\cdots$& $\infty$& $\infty^{\prime}$&$\cdots$ & $n^{\prime}$ & $(n-1)^{\prime}$ & $\cdots$  & $1^{\prime}$ & $0^{\prime}$ \\ \hline
$\psi_{0}$ & 0 & 0 & $\cdots$ & 0 & 0 & $\cdots$ & 0& 0&$\cdots$ & 0 & 0 & $\cdots$ & 0 & 0 \\
$\psi_{1}$ & 0 & 0 & $\cdots$ & 0 & 0 & $\cdots$ & 0& 0&$\cdots$ & 0 & 0 & $\cdots$ & 0 & 1 \\
$\vdots$ &  &  &  &  &  &  &  &  &  &  &  &  &  & \\
$\psi_{n}$ & 0 & 0 & $\cdots$ & 0 & 0 & $\cdots$ & 0& 0&$\cdots$ & 0 & 1 & $\cdots$ & 1 & 1 \\
$\vdots$ &  &  &  &  &  &  &  &  &  &  &  &  &  & \\
$\psi_{\infty}$ & 0 & 0 & $\cdots$ & 0 & 0 & $\cdots$ & 0& 0&$\cdots$ & 1 & 1 & $\cdots$ & 1 & 1\\ 
$\psi_{\infty^{\prime}}$ & 0 & 0 & $\cdots$ & 0 & 0 & $\cdots$ & 0& 1&$\cdots$ & 1 & 1 & $\cdots$ & 1 & 1\\ 
$\vdots$ &  &  &  &  &  &  &  &  &  &  &  &  &  & \\
$\psi_{n^{\prime}}$ & 0 & 0 & $\cdots$ & 0 & 1 & $\cdots$ & 1& 1&$\cdots$ & 1 & 1 & $\cdots$ & 1 & 1 \\
$\vdots$ &  &  &  &  &  &  &  &  &  &  &  &  &  & \\
$\psi_{1^{\prime}}$ & 0 & 1 & $\cdots$ & 1 & 1 & $\cdots$ & 1& 1&$\cdots$ & 1 & 1 & $\cdots$ & 1 & 1 \\
$\psi_{0^{\prime}}$ & 1 & 1 & $\cdots$ & 1 & 1 & $\cdots$ & 1& 1&$\cdots$ & 1 & 1 & $\cdots$ & 1 & 1 \\ \hline
\end{tabular}\end{center}
\caption{The functions $\psi$ in $\mathrm{C}(\Lambda^{\prime},2)$ and their values ($\psi(x)$ for $x\in\Lambda^\prime$)}
\label{TableFixedpoint2_p4}
\end{table}

Just as in $\Lambda$, both $\cdots 000$ and $ 111\cdots$ have an immediate neighbor in $\Lambda^{\prime}$. In addition, $\cdots 111$ and $ 000\cdots$ are mutual immediate neighbors in $\Lambda^{\prime}$.
\begin{defn}\label{DefOPP_p4}
Let $x$ be an infinite binary string of order type $\omega$ or $\omega^{*}$. Then $x^{\mathrm{opp}}$ is the infinite binary string which is generated from $x$ by first replacing every 0 with a 1 and vice versa, and then reversing the order type of the resulting string. For example, if $x=\cdots 00011$, and if the 0's and 1's in $x$ are replaced with 1's and 0's, respectively, the resulting string is $\cdots 11100$; reversing the order type of $\cdots 11100$ gives $ x^{\mathrm{opp}}=00111\cdots$.
\end{defn}
The relationship which consists of the following three conditions holds between $\Omega^{\prime}$ and $\Omega^{\prime(\mathrm{opp})}$ in $\Lambda^{\prime}$:\[\left(\vphantom{\Omega^{\prime(\mathrm{opp})}}\forall x\in\Omega^{\prime}\right)\,\left[x^{\mathrm{opp}}\in\Omega^{\prime(\mathrm{opp})}\right]\]
\[\left(\forall y\in\Omega^{\prime(\mathrm{opp})}\right)\,\left[y^{\mathrm{opp}}\in\Omega^{\prime}\right]\]
\[\left(\vphantom{\Omega^{\prime(\mathrm{opp})}}\forall x\in\Omega^{\prime}\right)\left(\forall y\in\Omega^{\prime(\mathrm{opp})}\right)\,\left[x\subseteq y^{\mathrm{opp}}\leftrightarrow x^{\mathrm{opp}}\supseteq y\right]\]
This relationship is an example of what is called adjunction in category theory~\cite{MacLane}.
Adjunction does not hold between $\Omega^{\prime}$ and $\Omega^{\mathrm{opp}}$ in $\Lambda=\Omega^{\prime}\cup\Omega^{\mathrm{opp}}$.  To see this, let $x=\cdots 111$, and note that $x\in\Omega^{\prime}$ but $x^{\mathrm{opp}}=000\cdots\,\not\in\Omega^{\mathrm{opp}}$.

\subsection{CPO of order type $\omega+1+\omega^{*}$; boundary element and adjunction}\label{SubSecBoundaryElement_p4}
Now we construct a new CPO $\hat{\Lambda}^{\prime}$ which is similar to $\Lambda^{\prime}$ but where the equivalents of the elements labeled with $\infty$ and $\infty^{\prime}$ are identical. 
\begin{prop}\label{PropProductOfTwoOrders_p4}
For $a\in\Omega^{\prime(\mathrm{opp})}$, let $\hat{\Omega}^{\prime}$ be the linear order with underlying set $\{a\}\times\Omega^{\prime}$ and ordering defined by\[(a,x_{1})\subseteq(a,x_{2})\leftrightarrow x_{1}\subseteq_{\,\Omega^{\prime}}x_{2}\]

 For $b\in\Omega^{\prime}$, let $\hat{\Omega}^{\prime(\mathrm{opp})}$ be the linear order with underlying set $\Omega^{\prime(\mathrm{opp})}\times\{b\}$ and ordering defined by\[(y_{1},b)\subseteq(y_{2},b)\leftrightarrow y_{1}\subseteq_{\,\Omega^{\prime\,(\mathrm{opp})}}y_{2}\]
Thus
\begin{equation}
\nonumber
\begin{array}{ccl}
&&\hat{\Omega}^{\prime}=\{(a,\,\cdots 000)\subseteq(a,\,\cdots 001)\subseteq(a,\,\cdots 0011)\subseteq\cdots\subseteq(a,\,\cdots 111)\}\\
&&\hat{\Omega}^{\prime(\mathrm{opp})}=\{(000\cdots,\, b)\subseteq\cdots\subseteq(0011\cdots,\, b)\subseteq(011\cdots,\, b)\subseteq(111\cdots,\, b)\}\end{array}
\end{equation}
Then $\hat{\Omega}^{\prime}\simeq\Omega^{\prime}$ and $\hat{\Omega}^{\prime(\mathrm{opp})}\simeq\Omega^{\prime(\mathrm{opp})}$. Moreover, the linear order  $\hat{\Omega}^{\prime}\cup\hat{\Omega}^{\prime\,(\mathrm{opp})}$ that extends the orderings in $\hat{\Omega}^{\prime}$ and $\hat{\Omega}^{\prime(\mathrm{opp})}$ and satisfies $(a,\cdots 111)\subseteq(000\cdots,b)$ is a CPO in which $(a,\cdots 111)$ is the sup of $\hat{\Omega}^{\prime}$ and $(000\cdots,b)$ is the inf of $\hat{\Omega}^{\prime\,(\mathrm{opp})}$.
\end{prop}

Let $\hat{\Lambda}^{\prime}$ be the CPO $\ \hat{\Omega}^{\prime}\cup\hat{\Omega}^{\prime(\mathrm{opp})}$ obtained from Proposition~\ref{PropProductOfTwoOrders_p4}      for $ a=000\cdots$ and $b=\cdots 111$, and let $m=(a,b)$ ($=(000\cdots,\,\cdots 111)$). Then $m$ is a boundary element of $\hat{\Lambda}^{\prime}$ in that $m\in\hat{\Omega}^{\prime}\,\cap\,\hat{\Omega}^{\prime\,(\mathrm{opp})}$ and $m$ is both the sup of $\hat{\Omega}^{\prime}$ and the inf of $\hat{\Omega}^{\prime\,(\mathrm{opp})}$. Also, $m$ has no immediate neighbor in $\hat{\Lambda}^{\prime}$, and $\hat{\Lambda}^{\prime}$ has order type $\omega+1+\omega^{*}$, so $\hat{\Lambda}^{\prime}\simeq\Lambda=\Omega^{\prime}\cup\Omega^{\mathrm{opp}}$ (and $\hat{\Lambda}^{\prime}\simeq\mathrm{C}(\hat{\Lambda}^{\prime},2)$) but $\hat{\Lambda}^{\prime}\not\simeq\Lambda^{\prime}=\Omega^{\prime}\cup\Omega^{\prime\,(\mathrm{opp})}$.

\begin{defn}\label{xopp1_p4}Let $(x,y)$ be an ordered pair of infinite binary strings each of which is of order type $\omega$ or $\omega^{*}$ (though $x$ is not necessarily of the same order type as $y$). Then $(x,y)^{\mathrm{opp}}=(y^{\mathrm{opp}},x^{\mathrm{opp}})$. For example, if $x=\cdots 00011$ and $ y=11111\cdots$, then $ x^{\mathrm{opp}}=00111\cdots$ and $y^{\mathrm{opp}}=\cdots 00000$, so \[(x,y)^{\mathrm{opp}}=(y^{\mathrm{opp}},x^{\mathrm{opp}})=(\cdots 00000,\, 00111\cdots)\]
\end{defn}

Note that $m^{\mathrm{opp}}=(000\cdots,\,\cdots 111)^{\mathrm{opp}}=(000\cdots,\,\cdots 111)=m$, and that the following conditions are satisfied:\[\hat{\Omega}^{\prime}=\{(x,y)^{\mathrm{opp}}:(x,y)\in\hat{\Omega}^{\prime(\mathrm{opp})}\}\]
\[\hat{\Omega}^{\prime(\mathrm{opp})}=\{(x,y)^{\mathrm{opp}}:(x,y)\in\hat{\Omega}^{\prime}\}\]
Moreover, it can easily be shown that adjunction holds between $\hat{\Omega}^{\prime}$ and $\hat{\Omega}^{\prime(\mathrm{opp})}$ in $\hat{\Lambda}^{\prime{\vphantom{\int}}}$.

\subsection{CPO of order type $1+\omega^{*}+\omega+1$}\label{SubSecConstructionOfV_p4}
Now we construct a CPO $V$ with order type $1+\omega^{*}+\omega+1$, using the following counterpart of Proposition~\ref{PropProductOfTwoOrders_p4}: 
\begin{prop}\label{PropProductOfTwoOrders2_p4}
For $a\in\Omega^{\prime}$, let $\Xi$ be the CPO with underlying set $\{a\}\times\Omega^{\prime(\mathrm{opp})}$ and ordering defined by \[(a,y_{1})\subseteq(a,y_{2})\leftrightarrow y_{1}\subseteq_{\,\Omega^{\prime\,(\mathrm{opp})}}y_{2}\]
For $b\in\Omega^{\prime(\mathrm{opp})}$, let $\Xi^{\mathrm{opp}}$ be the CPO with underlying set $\Omega^{\prime}\times\{b\}$  and ordering defined by  \[(x_{1},b)\subseteq(x_{2},b)\leftrightarrow x_{1}\subseteq_{\,\Omega^{\prime}}x_{2}\]
Thus
\begin{equation}\nonumber
\begin{array}{ccl}
&&\Xi=\{(a,\, 000\cdots)\subseteq\cdots\subseteq(a,\, 0011\cdots)\subseteq(a,\, 011\cdots)\subseteq(a,\, 111\cdots)\}\\
&&\Xi^{\mathrm{opp}}=\{(\cdots 000,\, b)\subseteq(\cdots 001,\, b)\subseteq(\cdots 0011,\, b)\subseteq\cdots\subseteq(\cdots 111,\, b)\}\end{array}
\end{equation}
Then $\Xi\simeq\Omega^{\prime(\mathrm{opp})}$ and $\Xi^{\mathrm{opp}}\simeq\Omega^{\prime}$. Moreover, the linear order $\Xi\cup\Xi^{\mathrm{opp}}$ that extends the orderings in $\Xi$ and $\Xi^{\mathrm{opp}}$ and satisfies $(a,\,\cdots 111)\subseteq(\cdots 000,\, b)$ is a CPO in which $(a,\,\cdots 111)$ is the sup of $\Xi$ and $(\cdots 000,\, b)$ is the inf of $\Xi^{\mathrm{opp}}$.
\end{prop}

Let $V$ be the CPO $\ \Xi\cup\Xi^{\mathrm{opp}}$ obtained from Proposition~\ref{PropProductOfTwoOrders2_p4} for $a=\cdots 000$ and $ b=111\cdots$, and let $m^{\prime}=(a,b)$ ($=(\cdots 000,\, 111\cdots)$). We can label the elements of $V$ as follows:
\begin{equation}\label{EqNewAssignment_p4}
\begin{array}{rcc}
(\cdots 000,\, 000\cdots)&\leftrightarrow&-\infty\\
\vdots&&\\
(\cdots 000,\, 001\cdots)&\leftrightarrow&-2\\
(\cdots 000,\, 011\cdots)&\leftrightarrow&-1\\
m^{\prime}=(\cdots 000,\, 111\cdots)&\leftrightarrow& 0\\
(\cdots 001,\, 111\cdots)&\leftrightarrow&+1\\
(\cdots 011,\, 111\cdots)&\leftrightarrow&+2\\
\vdots&&\\
(\cdots 111,\, 111\cdots)&\leftrightarrow&+\infty\\
\end{array}
\end{equation}

\begin{figure}[h]
\begin{center}
\includegraphics[width=120mm,height=31.2mm]{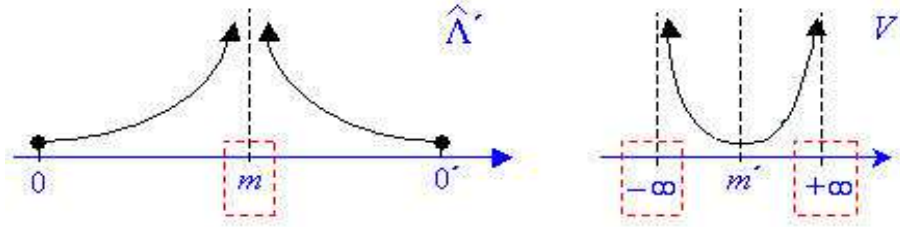}
\end{center}
\caption{CPO's $\hat{\Lambda}^{'}$ and $V$. The boxes indicate the elements that have no immediate neighbor.}
\label{FigLamdaAndV_p4}
\end{figure}

Then $m^{\prime}$ is a boundary element of $V$ in that $m^{\prime}\in\Xi\,\cap\,\Xi^{\mathrm{opp}}$ and $m^{\prime}$ is both the sup of $\Xi$ and the inf of $\Xi^{\mathrm{opp}}$. Also, $m^{\prime}$ has immediate neighbors in $V$, and $V$ has order type $1+\omega^{*}+\omega+1$. Therefore, $V$ is not isomorphic to any of the linear orders discussed earlier in this paper. (See Figure~\ref{FigLamdaAndV_p4} for a comparison of $\hat{\Lambda}^{\prime}$ and $V$.) Note that $m^{^\prime(\mathrm{opp})}=(\cdots 000,\, 111\cdots)^{\mathrm{opp}}=(\cdots 000,\, 111\cdots)=m^{\prime}$, hence that adjunction holds between $\Xi$ and $\Xi^{\mathrm{opp}}$ in $V$.

$V$ is not isomorphic to $\mathrm{C}(V,2)$. For suppose otherwise, and let $f:V\rightarrow\mathrm{C}(V,2)$ be an isomorphism. Then $f$ preserves the ordering of $V$, so $f$ maps the greatest element of $V$ (the element labeled with $+\infty$, which is $(\cdots 111,\, 111\cdots)$) to the greatest element of $\mathrm{C}(V,2)$, which is the all-1 function $\mathbf{1}$. Now define a function $\phi:V\rightarrow 2$ by\[\phi(x)=\left\{\begin{array}{lcl}0,&&x=(\cdots 000,\, 000\cdots)\\1,&&x\in V-\{(\cdots 000,\, 000\cdots)\}\end{array}\right.\]

 Note that $(\cdots 000,\, 000\cdots)$ is the least element of $V$ (i.e., the element labeled with $-\infty$), so $\phi$ is continuous (hence $\phi\in\mathrm{C}(V,2)$). Moreover, $\phi$ is the immediate predecessor of $\mathbf{1}$ in $\mathrm{C}(V,2)$, so $f^{-1}(\phi)$ must be the immediate predecessor of $f^{-1}(\mathbf{1})$ in $V$; however, $f^{-1}(\mathbf{1})=(\cdots 111,\, 111\cdots)$, which has no immediate predecessor.

\section{The finite binary strings in $S_{n}$ and their infinite extensions; the LR-transformation}\label{SecLR_p4}
In this section we introduce four ways to specify the finite binary strings that are elements of the finite CPO's $S_{n}$ ($n\ge 2$). Then we explore the extensions of those specifications to certain infinite binary strings that are of order type $\omega$ or $\omega^{*}$, and we define an operation (the  LR-transformation) on those infinite strings.
\subsection{Specifications of the strings in $S_{n}$}
For every $n\geq 2$, there are four ways to specify the elements of the finite linear order $(S_{n},\subseteq_{n})$ in (\ref{EqConstructionHigherOrderedSet_p4}).  Let $i,j\in\mathcal{N}$ such that $1\leq i\leq n$ and $1\leq j\leq n-1$.

{\bf Specification~I} $s_{i}^{(n)}=l_{1}\cdots l_{j}\cdots l_{n-1}$, where
\begin{equation}\label{SpecI_lj_p4}
l_{j}=\left\{\begin{array}{ccc}
0&\hspace*{.2in}&j\le n-i\\
1&\hspace*{.2in}&j>n-i\\
\end{array}\right.
\end{equation}

Then $s_{1}^{(n)}\subseteq_{n}s_{2}^{(n)}\subseteq_{n}\ldots\subseteq_{n}s_{n}^{(n)}$, where
\begin{equation}\label{SpecI_si_p4}
\begin{array}{ccc}
s_{1}^{(n)}&=&\underbrace{0\cdots 0}_{n-1}\\
s_{2}^{(n)}&=&\underbrace{0\cdots 0}_{n-2}1\\
s_{3}^{(n)}&=&\underbrace{0\cdots 0}_{n-3}11\\
&\vdots&\\
s_{n-1}^{(n)}&=&0\underbrace{1\cdots 1}_{n-2}\\
s_{n}^{(n)}&=&\underbrace{1\cdots 1}_{n-1}
\end{array}
\end{equation}

{\bf Specification~II} $s_{i}^{(n)}=\overline{l}_{1}\cdots\overline{l}_{j}\cdots\overline{l}_{n-1}$, where
\begin{equation}\label{SpecII_barlj_p4}
\overline{l}_{j}=\left\{\begin{array}{ccc}
0&\hspace*{.2in}&j<i\\
1&\hspace*{.2in}&j\ge i\\
\end{array}\right.
\end{equation}

Then $s_{1}^{(n)}\supseteq_{n}s_{2}^{(n)}\supseteq_{n}\ldots\supseteq_{n}s_{n}^{(n)}$, where
\begin{equation}\label{SpecII_si_p4}
\begin{array}{ccc}
s_{1}^{(n)}&=&\underbrace{1\cdots 1}_{n-1}\\
s_{2}^{(n)}&=&0\underbrace{1\cdots 1}_{n-2}\\
s_{3}^{(n)}&=&00\underbrace{1\cdots 1}_{n-3}\\
&\vdots&\\
s_{n-1}^{(n)}&=&\underbrace{0\cdots 0}_{n-2}1\\
s_{n}^{(n)}&=&\underbrace{0\cdots 0}_{n-1}
\end{array}
\end{equation}

{\bf Specification~III} $s_{i}^{(n)}=r_{n-1}\cdots r_{j}\cdots r_{1}$, where
\begin{equation}\label{SpecIII_rj_p4}
r_{j}=\left\{\begin{array}{ccc}
1&\hspace*{.2in}&j<i\\
0&\hspace*{.2in}&j\ge i\\
\end{array}\right.
\end{equation}

Then $s_{1}^{(n)}\subseteq_{n}s_{2}^{(n)}\subseteq_{n}\ldots\subseteq_{n}s_{n}^{(n)}$, where $s_{1}^{(n)},s_{2}^{(n)},\ldots,s_{n}^{(n)}$ are as in Specification~I, that is,
\begin{equation}\label{SpecIII_si_p4}
\begin{array}{ccc}
s_{1}^{(n)}&=&\underbrace{0\cdots 0}_{n-1}\\
s_{2}^{(n)}&=&\underbrace{0\cdots 0}_{n-2}1\\
s_{3}^{(n)}&=&\underbrace{0\cdots 0}_{n-3}11\\
&\vdots&\\
s_{n-1}^{(n)}&=&0\underbrace{1\cdots 1}_{n-2}\\
s_{n}^{(n)}&=&\underbrace{1\cdots 1}_{n-1}
\end{array}
\end{equation}

The only difference between (\ref{SpecIII_si_p4}) and (\ref{SpecI_si_p4}) is the order of indexation of the 0's and 1's in each element: from right to left in~(\ref{SpecIII_si_p4}), and from left to right in~(\ref{SpecI_si_p4}).

{\bf Specification~IV} $s_{i}^{(n)}=\overline{r}_{n-1}\cdots\overline{r}_{j}\cdots\overline{r}_{1}$, where
\begin{equation}\label{SpecIV_barrj_p4}
\overline{r}_{j}=\left\{\begin{array}{ccc}
1&\hspace*{.2in}&j\le n-i\\
0&\hspace*{.2in}&j>n-i\\
\end{array}\right.
\end{equation}

Then $s_{1}^{(n)}\supseteq_{n}s_{2}^{(n)}\supseteq_{n}\ldots\supseteq_{n}s_{n}^{(n)}$, where $s_{1}^{(n)},s_{2}^{(n)},\ldots,s_{n}^{(n)}$ are as in Specification~II, that is,
\begin{equation}\label{SpecIV_si_p4}
\begin{array}{ccc}
s_{1}^{(n)}&=&\underbrace{1\cdots 1}_{n-1}\\
s_{2}^{(n)}&=&0\underbrace{1\cdots 1}_{n-2}\\
s_{3}^{(n)}&=&00\underbrace{1\cdots 1}_{n-3}\\
&\vdots&\\
s_{n-1}^{(n)}&=&\underbrace{0\cdots 0}_{n-2}1\\
s_{n}^{(n)}&=&\underbrace{0\cdots 0}_{n-1}
\end{array}
\end{equation}

The only difference between (\ref{SpecIV_si_p4}) and (\ref{SpecII_si_p4}) is the order of indexation of the 0's and 1's in each element: from right to left in~(\ref{SpecIV_si_p4}), and from left to right in~(\ref{SpecII_si_p4}).

\subsection{Specifications of the corresponding infinite strings}\label{SubSecInfStrings_p4}
For every $n\geq 2$, all four specifications yield the same set of finite binary strings, $S_{n}$. When $ n\rightarrow\infty$, the four specifications yield four different sets of infinite binary strings, which we can define as follows:

{\bf Type~I} Let $i\ge 1$, and let $ t_{i}=\breve{l}_1\breve{l}_2\breve{l}_3\cdots$, where, for every $j\ge 1$,\[\breve{l}_j=\lim_{n\rightarrow\infty}l_{j}^{(n)}\] and $l_{j}^{(n)}$ is the $j$th bit (counting from the left) of the finite string $s_{i}^{(n)}$ in Specification~I.

Let $j\ge 1$, and recall that $l_{j}^{(n)}$ is defined for every $n$ such that $i\le n$ and $j\le n-1$, and that
\begin{equation}\label{TypeI_p4}
l_{j}^{(n)}=\left\{\begin{array}{ccc}
0&\hspace*{.2in}&j\le n-i\\
1&\hspace*{.2in}&j>n-i\\
\end{array}\right.
\end{equation}
The conditions $i\le n$ and $j\le n-1$ are equivalent to $n\ge i$ and $n\ge j+1$, respectively. Since $i,j\ge 1$, we have $j+i\ge i$ and $j+i\ge j+1$. Thus for every $n\ge j+i,\ \ l_{j}^{(n)}$ is defined; moreover, the condition $j\le n-i$ is satisfied, so $l_{j}^{(n)}=0$. From this it follows that $\displaystyle \breve{l}_j\mbox{\hspace{.05in}}(=\lim_{n\rightarrow\infty}l_{j}^{(n)})\mbox{\hspace{.05in}}=0$.

For every $ i\ge 1,\ \ t_{i}=000\cdots$; hence $(\{t_{i}:i\ge 1\},\subseteq)$ is the trivial linear order $\{\, 000\cdots\,\}$. Recall that $ 000\cdots$ is the inf of $\Omega^{\mathrm{opp}}$ in $\Lambda^{\prime}$.

{\bf Type~II} Let $i\ge 1$, and let $ t_{i}=\breve{\overline{l}}_1\breve{\overline{l}}_2\breve{\overline{l}}_3\cdots$, where, for every $j\ge 1$,\[\breve{\overline{l}}_j=\lim_{n\rightarrow\infty}\overline{l}_{j}^{(n)}\] and $\overline{l}_{j}^{(n)}$ is the $j$th bit (counting from the left) of the finite string $s_{i}^{(n)}$ in Specification~II.

Let $j\ge 1$. For every $n\ge j+i,\ \ \overline{l}_{j}^{(n)}$ is defined and
\begin{equation}\label{TypeII_p4}
\overline{l}_{j}^{(n)}=\left\{\begin{array}{ccc}
0&\hspace*{.2in}&j<i\\
1&\hspace*{.2in}&j\ge i\\
\end{array}\right.
\end{equation}
Thus
\begin{equation*}
\displaystyle \breve{\overline{l}}_j\mbox{\hspace{.05in}}(=\lim_{n\rightarrow\infty}l_{j}^{(n)})\mbox{\hspace{.05in}}=\left\{\begin{array}{ccc}
0&\hspace*{.2in}&j<i\\
1&\hspace*{.2in}&j\ge i\end{array}\right.
\end{equation*}
For every $i\ge 1,\ \ t_{i}$ is the infinite binary string of order type $\omega$ that (from left to right) consists of $(i-1)$ 0's followed by infinitely many 1's. In the finite case, the elements of $S_{n}$ are ordered as $s_{n}^{(n)}\subseteq\cdots\subseteq s_{1}^{(n)}$. Using the counterpart of that ordering scheme in the infinite case, we obtain $t_{i+1}\subseteq t_{i}$ for every $i\ge 1$, which yields the infinite linear order\[\{\,\cdots\,\subseteq\, 0011\cdots\,\subseteq\, 0111\cdots\,\subseteq\, 1111\cdots\,\}\]
This linear order, which has order type $\omega^{*}$, is $\Omega^{\mathrm{opp}}$.

{\bf Type~III} Let $i\ge 1$, and let $t_{i}=\cdots\breve{r}_3\breve{r}_2\breve{r}_1$, where, for every $j\ge 1$,\[\breve{r}_j=\lim_{n\rightarrow\infty}r_{j}^{(n)}\] and $r_{j}^{(n)}$ is the $j$th bit (counting from the right) of the finite string $s_{i}^{(n)}$ in Specification~III.

Let $j\ge 1$. For every $n\ge j+i,\ \ r_{j}^{(n)}$ is defined and
\begin{equation}\label{TypeIII_p4}
r_{j}^{(n)}=\left\{\begin{array}{ccc}
1&\hspace*{.2in}&j<i\\
0&\hspace*{.2in}&j\ge i\\
\end{array}\right.
\end{equation}
Thus 
\begin{equation*}
\displaystyle \breve{r}_j\mbox{\hspace{.05in}}(=\lim_{n\rightarrow\infty}r_{j}^{(n)})\mbox{\hspace{.05in}}=\left\{\begin{array}{ccc}
1&\hspace*{.2in}&j<i\\
0&\hspace*{.2in}&j\ge i\end{array}\right.
\end{equation*}
For every $i\ge 1,\ \ t_{i}$ is the infinite binary string of order type $\omega^{*}$ that (from right to left) consists of $(i-1)$ 1's followed by infinitely many 0's. In the finite case, the elements of $S_{n}$ are ordered as $s_{1}^{(n)}\subseteq\cdots\subseteq s_{n}^{(n)}$. Using the counterpart of that ordering scheme in the infinite case, we obtain $t_{i}\subseteq t_{i+1}$ for every $i\ge 1$, which yields the infinite linear order\[\{\,\cdots 0000\,\subseteq\,\cdots 0001\,\subseteq\,\cdots 0011\,\subseteq\,\cdots\,\}\]
This linear order, which has order type $\omega$, is $\Omega$.

{\bf Type~IV} Let $i\ge 1$, and let $t_{i}=\cdots\breve{\overline{r}}_3\breve{\overline{r}}_2\breve{\overline{r}}_1$, where, for every $j\ge 1$,\[\breve{\overline{r}}_j=\lim_{n\rightarrow\infty}\overline{r}_{j}^{(n)}\] and $\overline{r}_{j}^{(n)}$ is the $j$th bit (counting from the right) of the finite string $s_{i}^{(n)}$ in Specification~IV.

Let $j\ge 1$. For every $n\ge j+i,\ \ \overline{r}_{j}^{(n)}$ is defined and
\begin{equation}\label{TypeIV_p4}
\overline{r}_{j}^{(n)}=\left\{\begin{array}{ccc}
1&\hspace*{.2in}&j\le n-i\\
0&\hspace*{.2in}&j>n-i\\
\end{array}\right.
\end{equation}
For every $n\ge j+i$, the condition $j\le n-i$ is satisfied, so $\overline{r}_{j}^{(n)}=1$. From this it follows that $\displaystyle \breve{\overline{r}}_j\mbox{\hspace{.05in}}(=\lim_{n\rightarrow\infty}\overline{r}_{j}^{(n)})\mbox{\hspace{.05in}}=1$.

For every $i\ge 1,\ \ t_{i}=\cdots 111$; hence $(\{t_{i}:i\ge 1\},\subseteq)$ is the trivial linear order $\{\,\cdots 111\,\}$. Recall that $\cdots 111$ is the sup of $\Omega$ in both $\Lambda$ and $\Lambda^{\prime}$.

\subsection{LR-transformation}\label{SubSecLRTrans_p4}
There is a natural relationship between types~I and~IV, in that the sole infinite string $x$ of type~I  (namely, $ 000\cdots$) is $y^{\mathrm{opp}}$ for the sole infinite string $y$ of type~IV (namely, $\cdots 111$)---and, of course, $y=x^{\mathrm{opp}}$. Similarly, every infinite string $x$ of type~II  is $y^{\mathrm{opp}}$ for some infinite string $y$ of type~III, and every infinite string $y$ of type~III  is $x^{\mathrm{opp}}$ for some infinite string~$x$ of type~II.

\begin{figure}[h]
\begin{center}
\includegraphics[width=130mm,height=45mm]{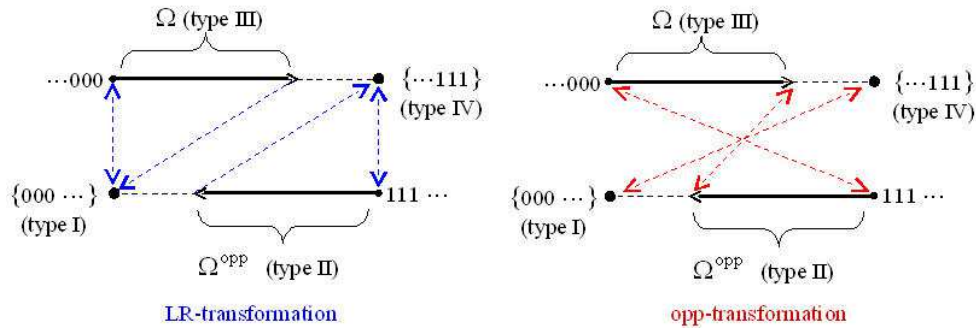}
\end{center}

\caption{LR-transformation and opp-transformation}
\label{FigOPP_LR_p4}\end{figure}

There is also a natural operation that transforms an infinite string of type~I  to an infinite string of type~III (and vice versa), and an infinite string of type~II  to an infinite string of type~IV  (and vice versa). For an infinite string $t_{i}$ of any of the four types, this operation consists of replacing $j$ with $n-j$ in the definition of $l_{j}^{(n)}$ in~(\ref{TypeI_p4}), $\overline{l}_{j}^{(n)}$ in~(\ref{TypeII_p4}), $r_{j}^{(n)}$ in~(\ref{TypeIII_p4}), or $\overline{r}_{j}^{(n)}$ in~(\ref{TypeIV_p4}), as appropriate, and reversing the order type of~$t_{i}$. We will refer to this operation as the LR-transformation, because it converts an infinite string whose bits are indexed starting from the left (L) to an infinite string whose bits are indexed starting from the right (R).  The LR- and opp-transformations for all four types of infinite strings are depicted in Fig.~\ref{FigOPP_LR_p4}.

For an ordered pair $(x,y)$ where $x$ and $y$ are infinite binary strings of one of the four types specified in section~\ref{SubSecInfStrings_p4} (though $x$ is not necessarily of the same type as $y$), define $(x,$\,$y)^{\mathrm{LR}}$ as $(x^{\mathrm{LR}},\, y^{\mathrm{LR}})$.

\section{Replication}\label{SecRep_p4}
Table~\ref{TableCPOs_p4} lists properties of the CPO's $\Lambda,\ \Lambda^{\prime},\ \hat{\Lambda}^{\prime}$, and $V$. 
\begin{table}[htbp]\begin{center}
\begin{tabular}{|l|cccc|}\hline
CPO& Adjunction & FPT & Boundary & Order type \\ \hline
$\Lambda$ & No & Applicable & N/A & $\omega+1+\omega^{*}$ \\ 
$\Lambda^{\prime}$ & Yes & Applicable & N/A & $\omega+1+1+\omega^{*}$ \\ 
$\hat{\Lambda}^{\prime}$ ($\simeq\Lambda$) & Yes & Applicable & $m$ & $\omega+1+\omega^{*}$  \\ 
$V$ ($\not\simeq\Lambda^{\prime}$)& Yes & Not applicable & $m^{\prime}$ & $1+\omega^{*}+\omega+1$ \\ \hline
\end{tabular}\end{center}
\caption{CPO's and their properties. FPT stands for ``Fixed Point Theorem.''}\label{TableCPOs_p4}
\end{table}

In continuum mechanics and thermodynamics, the basic notion of a body as a whole (i.e., Intent) and parts as infinitesimal body elements (i.e., Extent) has been discussed for several decades.  There arises the crucial problem of integration, i.e., of understanding how the body can glue the infinitesimal thermodynamical systems to obtain the global one. Owen proposes to approach this problem through the notion of sheaf~\cite{Owen1986}.

The theory of parts and a body naturally deals with sub-bodies (which must form a Boolean algebra)~\cite{Noll1986}.  Lawvere takes particular note of boundaries (which are not sub-bodies).  He points to a cartesian closed partially-ordered set as a convenient algebraic structure which includes these features (i.e., sub-bodies and boundaries)~\cite{Lawvere1986}, in which $\rightarrow$ is thought of as $\supseteq$ and hence cartesian product becomes $\cup$ while exponentiation becomes a binary operation akin to subtraction, which is characterized by
\[A\supseteq C\backslash B\hspace*{.1in}\Leftrightarrow\hspace*{.1in}A\cup B\supseteq C.\]
Then we can define
\[\sim A=1\backslash A\]
where 1 denotes the whole body; thus $\sim A$ is the smallest object such that $\sim A\cup A=1$.  And we can define the boundary of $A$ as
\[\partial A=A\cap\sim A.\]
Though Lawvere stated that the notion of boundary is just that of logical contradiction (within the realm of closed sets), how can we distinguish the boundary element $m$ of the CPO $\hat{\Lambda}^{\prime}$ (to which the FPT is applicable) from the one $m^{\prime}$ of the CPO $V$ (to which the FPT is not applicable)?  It might be consistent for $m$ to be a logical contradiction; however it is not for $m^{\prime}$.

$\hat{\Lambda}^{\prime}$ and $V$ are transformed into each other by the LR-transformation.  Since the FPT is applicable to $\hat{\Lambda}^{\prime}$ but inapplicable to $V$, they are different entities.  Therefore the LR-transformation transforms an entity into a different entity. For example, the infinite binary string $ 0111\cdots$ is transformed into a different infinite binary string $\cdots 111$ by the LR-transformation.  This means that $ 0111\cdots$ implies that it is defined with {\bf Type}~{\bf II} and $\cdots 111$ implies that it is defined with {\bf Type}~{\bf IV} (see section~\ref{SubSecInfStrings_p4}). That is, in this case, we can state:
\begin{itemize}
\item[(i)] Every existing infinite binary string remembers how it was defined and constructed.
\end{itemize}

On the contrary, the LR-transformation can be interpreted as replacing $j$ with $n-j$ in the definition of an`` existing'' infinite binary string (and then ``turning the string around''), that is, we may regard two different infinite binary strings which are transformed into each other by the LR-transformation as two different sides of the same coin.  (Therefore, if there are two CPO's such that the FPT is applicable to one but inapplicable to the other, and one of them is converted to the other by this interpretation of the LR-transformation, then we may interpret the conversion as an invalidation of the applicability of the FPT.)  That is, we can state that an observer who stands at the right endpoint observes $ 0111\cdots$ as $\cdots 111$.  Therefore, the string $ 0111\cdots$ itself does not imply that it is defined with {\bf Type}~{\bf II}.  The LR-transformation is interpreted as a transformation of the observer's positions (i.e., the left endpoint or the right endpoint of an infinite binary string), and therefore the LR-transformation acts as if it did nothing to an existing binary string.  (The counterpart of the LR-transformation in the finite strings in $S_{n}$ (see section~\ref{SecLR_p4}) is just an identity transformation.)  Thus we can introduce the concept $\pi\in\Pi$, corresponding to the existing infinite binary string.  So, in this case, we state: 
\begin{itemize}
\item[(ii)] No one, besides the string, knows how it was defined and constructed.
\end{itemize}

Though $(x,y)^{\mathrm{LR}}$ for such ordered pairs of infinite strings is well defined, the classification of such ordered pairs of infinite strings according to the four types is not. Consider $\hat{\Lambda}^{\prime}$, for example, which is of order type $\omega+1+\omega^{*}$. There is a natural isomorphism \[\varphi_{1}:\hat{\Lambda}^{\prime}\rightarrow\Omega\cup\Omega^{\prime\mathrm{opp}}=\Omega\cup\{\, 000\cdots\,\}\cup\Omega^{\mathrm{opp}}\] that pairs element $x$ of $\Omega$ with element $(000\cdots,\, x)$ of $\hat{\Lambda}^{\prime}$ (for every $ x\in\Omega$); pairs element $y$ of $\Omega^{\mathrm{opp}}$ with element $(y,\,\cdots 111)$ of $\hat{\Lambda}^{\prime}$ (for every $y\in\Omega^{\mathrm{opp}}$); and pairs $ 000\cdots$ with the boundary element $m=(000\cdots,\,\cdots 111)$ of $\hat{\Lambda}^{\prime}$. Now $\Omega$ is type III, $ 000\cdots$ is type I, and $\Omega^{\mathrm{opp}}$ is type II, so we could say that $\{(000\cdots,\, x):x\in\Omega\}$ is type III, $m$ is type I, and $\{(y,\,\cdots 111):y\in\Omega^{\mathrm{opp}}\}$ is type II. However, there is also a natural isomorphism \[\varphi_{2}:\hat{\Lambda}^{\prime}\rightarrow\Lambda=\Omega^{\prime}\cup\Omega^{\mathrm{opp}}=\Omega\cup\{\,\cdots 111\,\}\cup\Omega^{\mathrm{opp}}\] that pairs $\cdots 111$ with the boundary element $m$ (and is otherwise identical to~$\varphi_{1}$). Since $\cdots 111$ is type IV, we could just as well say that $m$ is type IV.

Now consider $V$, which has order type $1+\omega^{*}+\omega+1$. Thus \[V\simeq\Omega^{\prime\mathrm{opp}}\cup\Omega^{\prime}=\{000\cdots\}\cup\Omega^{\mathrm{opp}}\cup\Omega\cup\{\cdots 111\},\]which has the decomposition I + II + III + IV. However, there is no ``natural'' isomorphism in the sense of $\varphi_{1}$ or $\varphi_{2}$. Any isomorphism that pairs element $x$ of $\Omega^{\prime}$ with element $(x,\, 111\cdots)$ of $V$ (for every $x\in\Omega^{\prime}$) cannot also pair element $y$ of $\Omega^{\prime\mathrm{opp}}$ with element $(\cdots 000,\, y)$ of $V$ (for every $y\in\Omega^{\prime\mathrm{opp}}$): If $ y=111\cdots$, then $(\cdots 000,\, y)=m^{\prime}$, the boundary element of $V$, which would have been paired with element $x=\cdots 000$ of~$\Omega^{\prime}$ by such an isomorphism. Similarly, any isomorphism that pairs element $y$ of $\Omega^{\prime\mathrm{opp}}$ with element $(\cdots 000,\, y)$ of $V$ (for every $y\in\Omega^{\prime\mathrm{opp}}$) cannot also pair element $x$ of $\Omega^{\prime}$ with element $(x,\, 111\cdots)$ of $V$ (for every $x\in\Omega^{\prime}$). Thus there is no ``natural'' way to assign types to the elements of $V$.

Thus $\hat{\Lambda}^{\prime}$ and $V$ are transformed into each other by the LR-transformation; however the classification of $V$ according to the four types is not well defined.  $\hat{\Lambda}^{\prime}$ is ``naturally'' isomorphic to a linear order that can be decomposed as either III + I + II (corresponding to $\varphi_{1}$) or III + IV + II  (corresponding to $\varphi_{2}$), while $V$ is isomorphic to a linear order that has the decomposition I + II + III + IV; however there is no ``natural'' isomorphism (in the same sense that $\varphi_{1}$ and $\varphi_{2}$ are natural).  Thus, every infinite binary string of $\hat{\Lambda}^{\prime}$ is interpreted by (i), while every infinite binary string of $V$ is interpreted by (ii).

Here is a crucial ambiguity where we can debate whether the LR-transformation $\hat{\Lambda}^{\prime}$ to $V$ is a transformation between different entities or whether it acts as if it does nothing to the existing binary string, and therefore whether or not we can introduce the concept $\pi\in\Pi$, corresponding to the existing infinite binary string.  However, we do not stick to the problem to determine which is right, as IM does not.  Recall that in Gunji's model of IM~\cite{GunjiHarunaSawa}, it was not essential to inquire whether the defective observer destroys the Extent or simply invalidates the presupposition that leads to the paradox.  In the same sense, we stated that both (\ref{Eq_FormBinaryString_p4}) and $\pi\in\Pi$ are not strict mathematical statements in section~\ref{SecIntro_p4}.

In order for $\hat{\Lambda}^{\prime}$ and $V$ to be transformed into each other by the interpretation (i) of the LR-transformation, the two exclusive decompositions ``III + I + II'' and ``III + IV + II'' have to be combined into one pseudo-decomposition ``III + IV + I + II''.  So, conversely, we rather consider $\Lambda^{\prime}$ and $V$ are transformed into each other, where we insert a comma at the center of a binary string to divide the string into two parts when $\Lambda^{\prime}$ is transformed into $V$, and also we ignore the comma and stand at either the left or right endpoint of the infinite binary string when $V$ is transformed into $\Lambda^{\prime}$.  We will refer to this operation (which inherits the properties (i) and (ii) of the LR-transformation) as the LCR-transformation.  That is, first, we copy the boundary element $m=(000\cdots,\,\cdots 111)$ of $\hat{\Lambda}^{\prime}$, and, second, we project it into $ 000\cdots$ and $\cdots 111$, both of which are elements of $\Lambda^{\prime}$ and may be interpreted as $\mathrm{Int}(m)$ and $\mathrm{Ext}(m)$, respectively:
\begin{equation}\label{EqRep1_p4}
m=(000\cdots,\,\cdots 111)\xrightarrow[\mathrm{Copy}]{}\left\{\begin{array}{ccc}
m&\hspace*{.2in}&\xrightarrow[\mathrm{Projection}]{\mathrm{Int}(-)}000\cdots\\
m&\hspace*{.2in}&\xrightarrow[\mathrm{Projection}]{\mathrm{Ext}(-)}\cdots 111\\
\end{array}\right.
\end{equation}
Since we cannot possess two identical elements in a linear order, the two processes that ``{\it copy and then project}'' have to segue as if they are just one process. We call it a {\it replication}.

Thus the four infinite CPO's (i.e., $\Lambda,\ \hat{\Lambda}^{\prime},\ \Lambda^{\prime}$, and $V$) constructed in this paper are linked as:
\begin{equation}\label{EqSummary_p4}
\Lambda\xleftrightarrow[\mathrm{Dualization}]{\sim}\hat{\Lambda}^{\prime}\xrightarrow[\mathrm{Replication}]{}\Lambda^{\prime}\xleftrightarrow[\mathrm{LCR}]{}V
\end{equation}
The first CPO, $\Lambda$, (which is the one that Scott constructed) is isomorphic to the second one, $\hat{\Lambda}^{\prime}$.  We call the construction of $\hat{\Lambda}^{\prime}$ from $\Lambda$ (which we discussed in subsection~\ref{SubSecNonStandardRep_p4}) a {\it dualization} in (\ref{EqSummary_p4}).  The third one, $\Lambda^{\prime}$, is transformed into the fourth one, $V$, by the LCR-transformation.  Here is one important point.  We could construct a {\it complementarity} between an isomorphism $\psi_1:V\rightarrow\Omega^{\prime\mathrm{opp}}\cup\Omega^\prime$ that pairs element $\cdots 000$ of $\Omega^\prime$ with $m^\prime$, and an isomorphism $\psi_2:V\rightarrow\Omega^{\prime\mathrm{opp}}\cup\Omega^\prime$ that pairs element $111\cdots$ of $\Omega^{\prime\mathrm{opp}}$ with $m^\prime$.  Usually, for a boundary element $b\in\Upsilon\cap\Upsilon^{(\mathrm{opp})}$, where $\Upsilon$ is a linear order, we deem it just to be 
\begin{equation}\label{EqAnd_p4}
b\in\Upsilon\hspace*{.1in}\mathrm{and}\hspace*{.1in}b\in\Upsilon^{(\mathrm{opp})}
\end{equation}
On the contrary, we condemned it to the complimentary situation:
\begin{equation}\label{EqEitherOr_p4}
\mathrm{either}\hspace*{.1in}b\in\Upsilon\hspace*{.1in}\mathrm{or}\hspace*{.1in}b\in\Upsilon^{(\mathrm{opp})}
\end{equation}
corresponding to either $\psi_{1}$ or $\psi_{2}$ (each of which is a ``natural'' isomorphism in the same sense that $\varphi_{1}$ and $\varphi_{2}$ are natural), and then we interpreted (\ref{EqEitherOr_p4}) as:
\begin{equation}\label{EqCopyOr_p4}
b_{1}\in\Upsilon\hspace*{.1in}\mathrm{and}\hspace*{.1in}b_{2}\in\Upsilon^{(\mathrm{opp})}
\end{equation}
Here is the place where we can find the concept of replication as the process from (\ref{EqEitherOr_p4}) to (\ref{EqCopyOr_p4}) (corresponding to the process from $\hat{\Lambda}^{\prime}$ to $\Lambda^{\prime}$) by an invalidation of the applicability of the FPT to the CPO, $\Lambda$; namely the process is kicked by the boundary $m$ and the identity and/or indistinguishability between $b_{1}$ and $b_{2}$ is maintained by the boundary $m^{\prime}$.


\end{document}